\documentclass[aps,pre,twocolumn,groupedaddress,showpacs,showkeys]{revtex4}
\usepackage{epsfig}
\usepackage{epsf}
\usepackage{graphicx}

\begin{document}

\newcommand{\be}{\begin{equation}}
\newcommand{\ee}{\end{equation}}
\newcommand{\bea}{\begin{eqnarray}}
\newcommand{\eea}{\end{eqnarray}}
\newcommand{\nnn}{\nonumber \\}
\newcommand{\ba}{\begin{array}}
\newcommand{\ea}{\end{array}}
\newcommand{\nl}{\hspace{0.5cm}}
\newcommand{\nota}{ \bf }

\newcommand{\grad}{{\bf\vec{\nabla}}}
\newcommand{\xvec}{{\bf\vec{x}}}
\newcommand{\px}{\partial_x}
\newcommand{\py}{\partial_y}
\newcommand{\pt}{\partial_t}

%\newcommand{\version}{\underline{ condmat_control.tex }}

% Control.0.13 - PRL_control.tex is based in Control.0.13
% PRL_control - Corrections befor submission. Version submitted
% condmat_control - Based in PRL_control. Version submitted to condmat

%\begin{widetext}
% put long equation here
%\end{widetext}

%\bibliographystyle{apsrev}
%\pagestyle{empty}
\pagestyle{myheadings}
%\pagenumbering{arabic}

%\markboth{\version}{\version}

\title{Feedback control of unstable cellular solidification fronts}

\author{A. J. Pons}
\email{a.pons-rivero@neu.edu}
\author{A. Karma}
\affiliation{Physics Department and Center for Interdisciplinary
Research on Complex Systems, Northeastern University, Boston,
Massachusetts 02115}

\author{S. Akamatsu}
\altaffiliation{Permanent address: INSP, CNRS UMR 8875, Universit\'es
Paris VI and Paris VII, 140 rue de Lourmel, 75015 Paris, France}
\author{M. Newey}
\author{A. Pomerance}
\author{H. Singer}
\altaffiliation{Permanent address: Permanent address: Laboratorium
f$\ddot{u}$r
Festk$\ddot{o}$rperphysik,
ETH, CH-8093, H$\ddot{o}$nggerberg,
Z$\ddot{u}$rich,
Switzerland}
\author{W. Losert}
\affiliation{Department of Physics, IPST and IREAP, University of Maryland College Park, USA}

\date{\today}

\begin{abstract}
We present a numerical and experimental study of
feedback control of unstable cellular patterns in directional
solidification (DS). The sample, a dilute binary alloy, solidifies in
a 2D geometry under a
control scheme which applies local heating close to the
cell tips which protrude ahead of the other. For the experiments, we use a real-time
image processing algorithm to track cell tips, coupled with a
movable laser spot array device, to heat locally. We show,
numerically and experimentally, that spacings well below
the threshold for a period-doubling instability can be stabilized.
As predicted by the numerical calculations, cellular arrays become
stable, and the spacing becomes uniform through feedback control
which is maintained with minimal heating.
\end{abstract}

\maketitle

The control of cellular microstructures in directional solidification
(DS) of dilute binary alloys is a subject of both industrial and
fundamental interest \cite{KurzFisher,karmacta}. DS is produced in the
presence of a thermal gradient, $G$, which moves at velocity
$V_p$.
Cellular microstructures arise from the morphological instability of a
planar front when the velocity of the thermal gradient is above
some threshold, $V_c$, that depends
on the gradient $G$ and the
alloy concentration. Once the planar front becomes unstable,
it restabilizes (ideally) into a periodic array of solid fingers
or ``cells''. A dynamic competition between solute diffusion, in the
liquid, and capillary effects, at the moving solid-liquid interface,
determines the typical cell size, but the local
wavelength or cell spacing, $\Lambda$, admits a wide range of stable
values. When the average cell spacing, $\Lambda_0$, is above some spacing threshold, $\Lambda_{c}$, the array is
stable and achieves a steady configuration. When $\Lambda_0 <
\Lambda_{c}$, some cells, generally those of larger local
wavelength, grow faster than their neighbors. This leads to the amplification of some
modes and, eventually, to the elimination of, approximately, one cell
out of two \cite{Warren1993, kopcz} (period-doubling instability).
During the initial evolution of the cell array, the elimination process is repeated, increasing
progressively $\Lambda_0$, until a stable configuration with $\Lambda_0
> \Lambda_c$ is reached.

In a previous experimental work, Lee and Losert \cite{Lee2004} (see also
\cite{Losert1998}) have shown, using the so called ``combing method'',
that it is possible to select a uniform
cell spacing within the stable range. The combing method consists
of using strong local
temperature perturbations in the vicinity of the front during
the initial transient evolution. The thermal perturbation sets the
periodicity to the cell array. However, a permanent control
of cellular patterns outside the stability domain had not yet been
achieved. In this letter, we propose a new scheme to achieve this
control and demonstrate its feasibility in both phase-field simulations
and experiments. Our
scheme works, essentially, by slowing down the growth of any cell which
overgrows the average position of other cells
in the direction of growth $y$. For this
purpose, we apply local heating close to
the protruding cell tips. A key feature of this scheme
is that the amplitude of feedback perturbations (i.e. the magnitude
of heating) essentially vanishes in the
controlled state.

We have performed numerical calculations
using a modified version of a recently proposed quantitative
phase-field model of binary alloy solidification
\cite{Karma1996,Karma1998,Karma2001,Echebarria2004}. For the experiments,
we use a model transparent alloy, namely, succinonitrile
(SCN)-coumarin 152 (C152), in thin sample (see, e.g., \cite{Lee2004}
and refs. therein).

In the numerical calculations,
we use a feedback control scheme which
is a simple step
function, i.e. the amount of heat injected in a spot-like
region is constant regardless of how far the targeted cell grows
beyond the average cell position $\bar{y}$. We consider an array of
$N$ cells in a rigid box (with no-flux boundary conditions). For a given
cell $q$, local heating at the tip is applied when the distance from
the average cell position, $y_q-\bar{y}$
is larger than a pre-defined cut-off $\delta > 0$. The temperature
field $T(\xvec,t)$, neglecting the production of
latent heat (``frozen temperature approximation''),
can be expressed as follows:

\be
T(\xvec,t)=T_0+G(y-V_pt)+p(\xvec,t),\label{Eq:1}
\ee
%\begin{widetext}
\be
p(\xvec,t)=\sum_q g\Lambda_0~ H\left[y_q(t)-\bar{y}(t)-\delta\right]
\exp\left(\frac{-(\xvec-\xvec_q(t))^2}{\xi^2}\right),\label{Eq:2}
\ee
%\end{widetext}
where $\xvec$ is the two-dimensional vector position,
$p(\xvec,t)$ is the imposed thermal perturbation, $H[...]$ is
the Heaviside step function, and the sum extends to the $N$ cells.
These equations enter as a modification of
Eqs. (132-133) in Ref. \cite{Echebarria2004}. They approximate
the fact that each heating spot in the experiments
results in the build up of a Gaussian
bump in the thermal field of constant width, $\xi$, and amplitude,
$g\Lambda_0$. A precise
determination of the phase diagram for the SCN-C152 alloy used in
this study is presently lacking. Therefore,
for sake of realism, we carry out the phase-field
simulations for physical parameters
(Table \ref{Ta:1}) estimated in Ref. \cite{Georgelin1998} for a SCN-X
alloy (where X stems for an unknown impurity).

\begin{table}[t]
\begin{center}
\begin{tabular}{l r} \hline \hline
$|m|c_{\infty}$ (shift in melting temperature) & $2~K$ \\
$D$ (diffusion coefficient) & $10^{-9}~m^2/s$ \\
$\Gamma$ (Gibbs-Thompson coefficient) & $6.48 \times 10^{-8}~Km$ \\
$V_p$ (pulling speed) & $ 32 ~\mu m/s$\\
$G$ (thermal gradient) & $143.587 ~K/cm$\\
$d_0$ (capillary length) & $1.3 \times 10^{-2}~\mu m$\\
$k$ (partition coefficient) & 0.3\\
$\epsilon_4$ (0.7 \% anisotropy) & 0.007 \\
\hline \hline
\end{tabular}
\caption{\small Parameters for impure succinonitrile
used in phase-field simulations \cite{Georgelin1998}. See also
\cite{Echebarria2004}.}\label{Ta:1}
\end{center}
\end{table}

\begin{figure}
\rotatebox{-90}{\includegraphics[width=5.6cm]{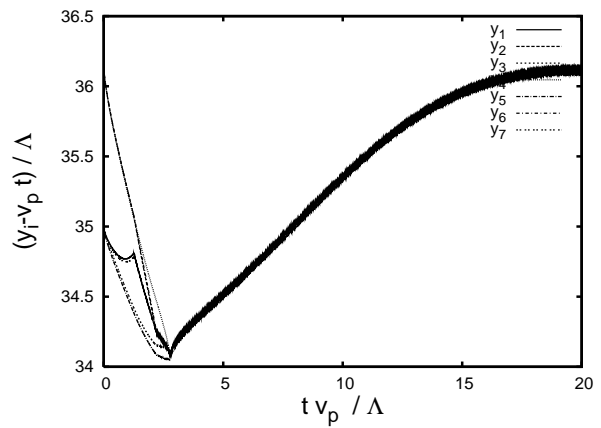}}
\caption{ Evolution in time of the tip positions, $y_q(t)$, in the growth
direction referred to a given isotherm with control on.
$g/G$=2.5, $\delta/\Lambda=0.5 $\% and, $\xi/\Lambda$=0.97.
$\Lambda$=11.3 $\mu m$. \label{Fi:1} }
\end{figure}

\begin{figure}
\rotatebox{-0}{\includegraphics[width=6cm]{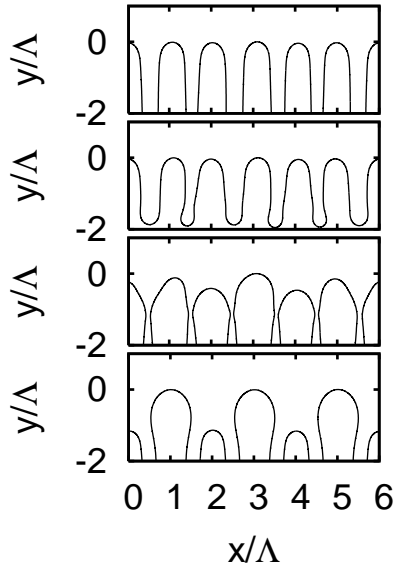}}
\caption{Snapshots of the front evolution of the system shown in Figure
\ref{Fi:1}. The near cell-tip region is displayed for
simplicity. Times (increasing upwards): (a) 0.0 s, (b) 0.6952 s, (c)
1.3992 s and (d) 2.1032 s.
 \label{Fi:2} }
\end{figure}

The feasibility of our control scheme is best illustrated
by comparing the dynamic evolution of a strongly unstable
cellular array with and without control.
The wavelength of the array $\Lambda_0=11.3~\mu
m$ is well below the stability threshold $\Lambda_{c}\approx
50 ~\mu m$ for cell elimination \cite{EchebarriaUnpublished}.
Without control,
the known spatial period-doubling instability
that leads to cell elimination is observed. In contrast, with
control on, this instability is suppressed. This is
illustrated in Figure \ref{Fi:1} where we
show the evolution in time of the front position, $y_q$, at specific $x$ positions
that initially correspond to cell tips, when control is on. It is observed that $\bar{y}$ decreases
at the same time that $y_q$ dispersion
decreases. Later, when $y_q$ dispersion is reduced, the
average position advances arriving to a
steady state value.
In Figure \ref{Fi:2} we show some snapshots of the front
evolution. Due to
the large initial value of $|y_q-\bar{y}|$
along the pattern, a relatively
strong (or frequent) heating is necessary for melting protruding cells
backwards (decreasing $\bar{y}$). This entails a transitory disorder visible in
Fig. \ref{Fi:2}b. However, this large-perturbation stage ceases as
soon as cell tips reach an almost equal undercooling
(Fig. \ref{Fi:2}c). Eventually, a uniform pattern is
stabilized
(Fig. \ref{Fi:2}d). We will see below
that a uniform $\Lambda(x)$ distribution,
obtained numerically by construction of the array,
will also be obtained in the experiments.
Interestingly, the heating frequency (not shown)
decreases during the whole process, and reaches a very small value
when control has been achieved. Note also that
the system could not be
controlled by setting $\delta =0$ ($\delta/\Lambda=0.5$\% in
Fig. \ref{Fi:2}). The reason for this is that it is practically
impossible to maintain all tips at exactly the average fixed position.
Therefore, our feedback algorithm with $\delta =0$ heats continuously
some of the tips and leads to a
completely melt out of cells, destroying the periodicity of the
array. Naturally, feedback control also
fails for
$\delta$ values larger than, say, $0.1~\Lambda$. Note finally
that
a similar cut-off (about one pixel) was also introduced in our
experimental feedback program.

In the experiments, we solidified a
binary SCN-0.1wt\% C152 alloy
in a gradient $G = 10 ~Kcm^{-1}$
, and with $V_p= 2 - 20 ~\mu
\rm{ms^{-1}}$ (same device as in Ref. \cite{Lee2004}). The alloy, which crystallizes in a $bcc$ cubic crystal,
is confined in a 100 $\mu m$ thick, 2 $mm$ wide, 150 $mm$ long,
glass-wall microtube. Prior to DS experiments, a single crystal with
its [100] axis almost parallel,
within less than $2\rm{^o}$, to the solidification axis was selected
by an empirical procedure and grown in such a way that it fills
the container; it served as a permanent seed for further
experiments. Experiments justify the use of a $0.7\%$ anisotropy for
the surface tension \cite{anisotropy}. The local heating system is a
holographic laser tweezer system (BioRyx200 from Arryx Inc) which
consists of a 2-W NdYAG ($\lambda$=532~nm) laser that is focused onto
a spatial light modulator. The single laser spot is then split into a
multiple diffraction spot pattern, which is projected into the imaged
region through a (4x) objective of an inverted microscope. The system
software allows independent positioning of hundreds of spots
into controlled arrangements.
Heating in the liquid is due to
partial absorption of light by the fluorescent dye \cite{Williams93}.
In a first
approximation, a laser spot (about 10 $\mu m$ in diameter) acts as a
point-like source of heat which diffuses rapidly in a quasi-bulk
medium, including the thick glass walls. A Gaussian bump superimposed
to the linear gradient (see Eq. (\ref{Eq:2})) is, thus, a realistic
representation of the modified thermal field. From a rough
calibration, we estimate that for short-exposure times (of order 1~s)
and low (about 0.2~$W$) laser power, the material is heated by less
than 0.1 Kelvin in a region which extends over less than $40~\mu
\rm{m}$. The numerical simulation heating power corresponds to
$g\Lambda_0 \approx 0.4$ Kelvin and extends in $\xi \approx$ 10 $\mu m$.
There is no measurable overlapping between two neighboring laser spots, which are much smaller than $\Lambda_0$ (typically $100~\mu \rm{m}$).

We image the solidification front with a (1024x1280 pixel$^2$) digital
camera.
After some image processing, the front line is detected and smoothed
with a moving central average of 5 pixels and the list of tips is detected.
They are subsequently sent to laser controller after calculating the
deviation from the average tip position.
We use a feedback control program based on the numerical
simulations to automatically place laser spots in the liquid region
ahead of the protruding
tips. By restricting the
number of controlled spots to eight ($N=8$), we are able to update the
laser spots at approximately one Hz.

\begin{figure}
\rotatebox{-0}{\includegraphics[width=5.5cm]{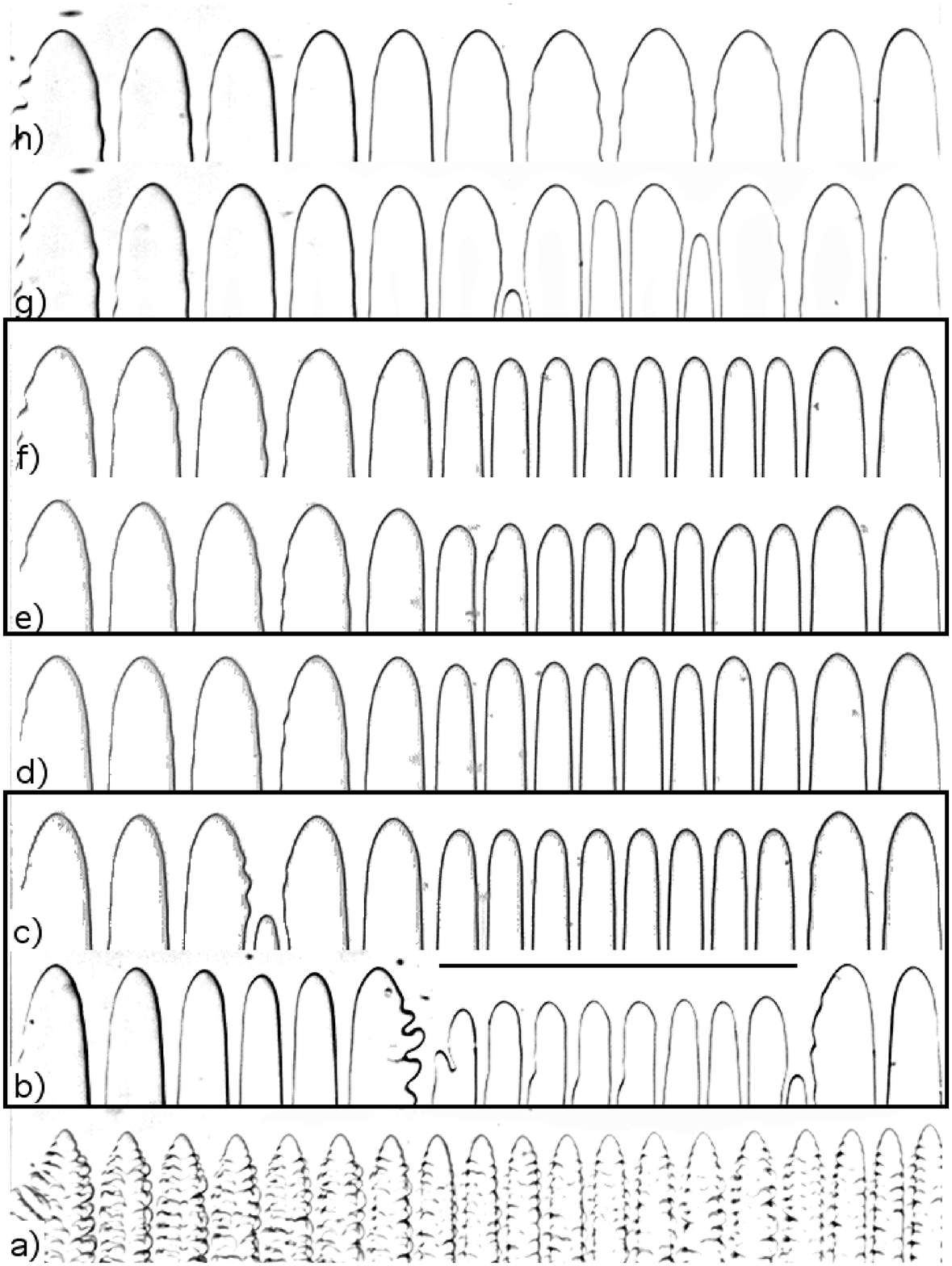}}
\caption{Experimental sequence of images during feedback control
of a small-spacing cell front pattern (time increases upwards). Also see Fig. \ref{Fi:4}. (a) Initial, high-velocity
cell pattern ($t=0$; $V_p= 18~\mu \rm{ms^{-1}}$) obtained with the combing
method (see text); (b) Transient, distorted, pattern due to strong perturbations just after
control is turned on ($t=1100$ s; $V= 6.1~ \mu \rm{ms^{-1}}$); (c) Stabilized pattern ($t=1520$ s; $V_p= 4.6~\mu \rm{ms^{-1}}$); (d) Period-doubling instability after control has been turned off ($t=1610$ s); (e)
Distorted pattern after
control is turned on again ($t=1620$ s); (f) Re-stabilized pattern
($t=2100$ s); (g) Highly unstable pattern (control off) ($t=2300$ s); (h) Large-spacing (uncontrolled) pattern. Horizontal bar: controlled region. Frames: feedback control is on. Horizontal dimension:
2000 $\mu m$.
 \label{Fi:3} }
\end{figure}

\begin{figure}
\rotatebox{-0}{\includegraphics[width=7cm]{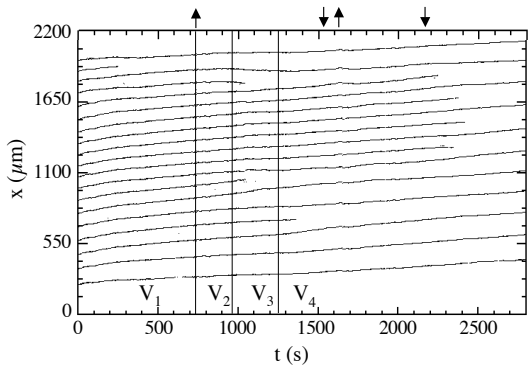}}
\caption{Lateral cell tip position $x$ as a function of time (same run as Figure
\ref{Fi:3}). Vertical bars: velocity jumps ($V_1 = 18.4~ \mu \rm{ms^{-1}}$;
$V_2 = 9.2~ \mu \rm{ms^{-1}}$; $V_3 = 6.1~ \mu \rm{ms^{-1}}$; $V_4 = 4.6~
\mu \rm{ms^{-1}}$). Upward (downward) arrows: feedback control on (off).
Lateral drift of the pattern is due to a
slight misalignment of the [100] axis of the single crystal with axis
$y$.
 \label{Fi:4}}
\end{figure}

A sequence of images during a typical feedback control experiment is shown
in Figure \ref{Fi:3}. A simplified spatio-temporal
diagram of the experiment is shown in Fig. \ref{Fi:4}. First, we perform a fast partial
melting of the sample to obtain a planar solid-liquid interface at rest (not shown). Then, we
start pulling at relatively high velocity ($V_1 \approx 18~\mu
\rm{ms^{-1}}$) which allows us to obtain a cellular array
with small spacing by using the combing method (Fig. \ref{Fi:3}a). Then, we switched off
the combing laser array, turned on control, and decreased the
velocity stepwise down to a final value $V_4= 4.6~\mu \rm{ms^{-1}}$
($\approx V_1/4$) at which the pattern is unstable. After a relatively short transient, during which the pattern is strongly perturbed due to frequent illumination (Fig. \ref{Fi:3}b), a well controlled, small-spacing pattern is eventually obtained (Fig. \ref{Fi:3}c). Note that the area accessible to
the laser spot array is limited to about 1~mm, thus, only one half of
the cellular pattern may be controlled. The other half was imaged but
grew freely without control. As it can be seen in the spatio-temporal
diagram of Fig. \ref{Fi:4}, many cells are eliminated
outside the controlled window, while the number of controlled
cells remains constant in the
controlled area.

We then switched intentionally the laser light off (after about 10 minutes of successful control) during a few minutes, 
and observed the onset of the period-doubling instability (Fig. \ref{Fi:3}d). By turning feedback control on again, we were 
able to prevent cell elimination and to restabilize a small-spacing cell
pattern similar to the initial one (Fig. \ref{Fi:3}f). We
stress the striking resemblance between numerical (Fig. \ref{Fi:2})
and experimental (Figs. \ref{Fi:3}d to \ref{Fi:3}f) runs --including the transient
stage after the laser is turned on (Fig. \ref{Fi:2}b and Fig. \ref{Fi:3}e). Finally, we switched feedback control off again and let the instability fully develop, and, as expected, approximately one cell out of two is eliminated in the previously controlled area (Figs. \ref{Fi:3}g and \ref{Fi:3}h). Let
us make two important remarks.

In numerical as well as in experimental runs, the spacing distribution
in a well controlled pattern is remarkably uniform (it is not so
outside the controlled area) as seen, for example, in Figure \ref{Fi:3}b. In addition, the controlled
small-spacing cells sit slightly behind, thus have a larger undercooling than
the large uncontrolled ones, as expected. This is not due to the added heat
from the feedback control, but due to interactions between cells.

The second remark concerns the heating power applied to the cell
pattern, or, equivalently,
the laser spot exposure frequency $f$ ahead
of cells. Our main observation is that $f$ gradually decreases as
control goes on. In an experiment performed at $V_p=2.9~\mu
\rm{ms^{-1}}$ (not shown), we measured $f \approx 0.5~s^{-1}$ at the
beginning of the run, and, $f \approx 0.05~s^{-1}$, 10 minutes later.
As expected, the final characteristic time $1/f \approx 20~s$ is smaller than the
characteristic
amplification time of the period-doubling instability ($\tau\approx $
80 s)
measured in situ in an uncontrolled, unstable pattern.
Furthermore, we always
performed feedback control sequences much longer (more than
10 minutes) than $\tau$. It is interesting to note, also, that the final
$1/f$ is comparable
to the so-called diffusion time $\tau_d = D/V^2 = 10 - 100 s$ ($D$
falls in the $10^{-10}-10^{-9}~m^2s^{-1}$ range), which signals a
quasi steady regime. Therefore, control can be
maintained as long as wanted.

In conclusion, we have shown experimentally and numerically that
highly unstable cellular solidification arrays
can be stabilized using
feedback control. Experiment and simulation
showed similar dynamical evolution as the array was
controlled.
Unlike feedback stabilization schemes for planar fronts slightly above
$V_c$
proposed by Savina et al,~\cite{Savina},
our approach stabilizes efficiently
highly unstable states and can be implemented
experimentally with only
a discrete number of controllable heating points.
An alternate control scheme which includes a tunable heating power for
each spot has been implemented
numerically \cite{PonsUnpublished}; an experimental realization is
currently in progress. Although
a quantitative comparison between
numerical and experimental results is not yet
possible because of the lack of detailed knowledge of the alloy
phase diagram, the
observed qualitative behaviors are identical.

The use of other localized physical perturbations
generated e.g. by x-rays or ultrasound
could potentially make it possible to extend this control scheme to
metallic alloys where the growth of much finer array structures
with superior mechanical properties is of considerable practical interest.
In addition, the ability to access experimentally unstable
steady-state patterns provides an important new tool to enrich
our fundamental understanding of
pattern formation in directional solidification.

We thank B. Echebarria for helpful discussions. This research was
supported by NASA grant NNM04AA15G
and by Ministerio de Educaci\'on y Ciencia under the
grant number EX2005-0085, and by a Research Corporation Research
Innovation Award.

\end{document}